\documentclass[12pt,preprint]{emulateapj}

\usepackage{graphicx}
\usepackage{color}

\newcommand{\beq}{\begin{equation}}
\newcommand{\eeq}{\end{equation}}
\newcommand{\beqa}{\begin{eqnarray}}
\newcommand{\eeqa}{\end{eqnarray}}

\begin{document}

\title{\bf A possible evidence of the gluon condensation effect in cosmic positron and gamma-ray spectra\\
}

\author{Wei Zhu$^1$\altaffilmark{*}, Peng Liu$^1$, Jianhong Ruan$^1$ and Fan Wang$^2$}

\affiliation{$^1$Department of Physics, East China Norma University,
Shanghai 200241, China\\
$^2$Department of Physics, Nanjing University, Nanjing,210093, China
} \altaffiltext{*}{ Corresponding author: wzhu@phy.ecnu.edu.cn}

\begin{abstract}
     The gluon condensation effect in cosmic proton-proton collisions at high energy
is used to explain an excess in the positron spectrum observed by
AMS. We find that this excess may originate from the GC-effect in
Tycho's supernova remnant.
\end{abstract}


\setlength{\parindent}{.25in}

\section{Introduction}

    The precise measurement of the positron flux in cosmic
rays (CRs) with the Alpha Magnetic Spectrometer (AMS) on the
International Space Station exhibits complex energy dependence
(Aguilar et al.2019). The results show that a significant excess is
added on a diffuse background around $300\sim 400~GeV$ of the
positron energy. A new extra source is predominantly suggested as
dark matter annihilation or other astrophysical sources. A following
key point is where is this local source? We can't "see" dark matter.
We also can't track the positron because of their complex
trajectories in interstellar space. While if we know this extra
source radiates a characteristic gamma-ray spectrum, which is
closely related to the excessive positrons, then we can find the
extra source. The GC mechanism provides this possibility.

   Gluons inside proton dominate the proton collisions at high energy
and their distributions obey the evolution equations based on
Quantum Chromodynamics (QCD). QCD analysis shows that the evolution
equations will become nonlinear due to the initial gluons
correlations at high energy and these will result in the chaotic
solution beginning at a threshold energy (Zhu et al, 2008; Zhu et
al, 2006). Most surprisingly, the dramatic chaotic oscillations
produce strong shadowing and antishadowing effects, they converge
gluons to a state at a critical momentum (Zhu et al, 2017). This is
the gluon condensation (GC) in proton. The GC should induce
significant effects in the proton collision processes, provided the
collision energy is higher than the GC-threshold $E_{p-p}^{GC}$. The
GC model was be used to explain the excess in the fluxes of cosmic
gamma-ray and electron-positron (Feng et al. 2018; Zhu et al. 2018).

    In this work we will present a new evidence, which shows that
the positron excess observed by AMS originates from the GC-effect in
Tycho's supernova remnant. Our idea is straightforward. High energy
gamma-ray and electron-positron fluxes in CRs may originate from
leptonic processes (bremsstrahlung, inverse Compton scattering and
electromagnetic acceleration mechanism), they are also the products
in hadronic processes, where $p+p\rightarrow \pi^0\rightarrow
2\gamma$, and through strong electromagnetic field inside source
$\gamma\rightarrow e^++e^-$. We take hadronic framework but add the
GC-effect in this work.  When a large amount of gluons condense in a
critical momentum space, it must inevitably increase the number of
secondary particles ($\pi, \gamma, e^+ ~and ~e^-$) at the
corresponding energy threshold and breaks the power-law. In
particular, the positron excess always accompanied by a special
broken gamma-ray spectrum since a considerable part of these two
fluxes (see Equations (2.10) and (2.13)) are interrelated. It
implies that the gamma-ray spectrum corresponding to the excess
positron flux has its characteristic shape. Thus, we can judge which
source produces the AMS positron excess using a directly observable
gamma-ray spectrum.

    We will give the relating formulas of the GC model for
the explanation of gamma-ray and positron flux in Section 2. A
detailed description can be found in (Feng et al.2018; Zhu et
al.2018). In Section 3 we analyze the Tycho's gamma-ray spectrum and
compare it with the AMS positron flux. We find that their
consistency with the theoretical predictions reaches
$\chi^2/d.o.f.=0.54$ and $0.45$, respectively. This result strongly
suggests that the positron excess observed by AMS mainly origins
from nearby Tycho's supernova remnant. The discussions and summary
are given in Section 4

\section{The GC model}

     High energy gamma ray and electron-positron fluxes in CRs may
origin from the following processes $p+p\rightarrow \pi^0\rightarrow
2\gamma$, and $\gamma\rightarrow e^++e^-$. The number of secondary
pions in proton-proton collision is determined by the gluon
distribution inside proton (Field et al.1977). Usually, these pions
have a small kinetic energy (or low momentum) at the center-of-mass
(C.M.) system and form the central region in the rapidity
distribution. Gluons in protons may condensed at a critical momentum
($x_c, k_c$). One can image that this GC effect should be appeared
in CR spectra. A large number of gluons at the central region due to
GC effects create the maximum number $N_{\pi}$ of pions, which take
up all available kinetic energy, where we neglect other secondary
particles. Using general relativistic invariant and energy
conservation, we have

$$(2m_p^2+2E_{p-p}m_p)^{1/2}=E^*_{p1}+E^*_{p2}+N_{\pi}m_{\pi}, \eqno(2.1)$$
$$E_{p-p}+m_p=m_p\gamma_1+m_p\gamma_2+N_{\pi}m_{\pi}\gamma, \eqno(2.2)$$where
$E^*_{p_i}$ is the energy of leading proton at the C.M. system,
$\gamma_i$ are the corresponding Lorentz factors. Using the
inelasticity $K$, we set

$$E^*_{p1}+E^*_{p2}=(\frac{1}{K}-1)N_{\pi}m_{\pi},  \eqno(2.3)$$and

$$m_p\gamma_1+m_p\gamma_2=(\frac{1}{K}-1) N_{\pi}m_{\pi}\gamma. \eqno(2.4)$$
One can easily get the solutions $N_{\pi}(E_{p-p},E_{\pi})$ for the
$p-p$ collisions

$$\ln N_{\pi}=0.5\ln E_{p-p}+a, ~~\ln N_{\pi}=\ln E_{\pi}+b,  \eqno(2.5)$$
$$~~ where~E_{\pi}
\in [E_{\pi}^{GC},E_{\pi}^{max}].$$ The parameters in the
GeV unit read

$$a\equiv 0.5\ln (2m_p)-\ln m_{\pi}+\ln K, \eqno(2.6)$$ and

$$b\equiv \ln (2m_p)-2\ln m_{\pi}+\ln K. \eqno(2.7)$$Equation (2.5) gives the relations among $N_{\pi}$, $E_{p-p}$ and
$E_{\pi}^{GC}$ by one-to-one, it leads to a following
GC-characteristic spectrum.

    According to the above mentioned hadronic framework of gamma-ray
spectrum, we write

$$\Phi^{GC}_{\gamma}(E_{\gamma})=C_{\gamma}\left(\frac{E_{\gamma}}{E_{\pi}^{GC}}\right)^{-\beta_{\gamma}}
\int_{E_{\pi}^{min}}^{E_{\pi}^{max}}dE_{\pi}
\left(\frac{E_{p-p}}{E_{p-p}^{GC}}\right)^{-\beta_p}$$
$$\times N_{\pi}(E_{p-p},E_{\pi})
\frac{d\omega_{\pi-\gamma}(E_{\pi},E_{\gamma})}{dE_{\gamma}},
\eqno(2.8)$$ where  the GC-effect enters $\Phi^{GC}_{\gamma}$ via
Eqution (2.5). The spectral indexes $\beta_{\gamma}$ and $\beta_p$
denote the propagating loss of gamma-rays and protons, respectively;
$C_{\gamma}$ incorporates the kinematic factor with the flux
dimension and the percentage of $\pi^0\rightarrow 2\gamma$. The
normalized spectrum for $\pi^0\rightarrow 2\gamma$ is

$$\frac{d\omega_{\pi-\gamma}(E_{\pi},E_{\gamma})}{dE_{\gamma}}
=\frac{2}{\beta_{\pi}
E_{\pi}}H[E_{\gamma};\frac{1}{2}E_{\pi}(1-\beta_{\pi}),
\frac{1}{2}E_{\pi}(1+\beta_{\pi})], \eqno(2.9)$$wnere $H(x;a,b)=1$
if $a\leq x\leq b$, and $H(x;a,b)=0$ otherwise, $\beta_{\pi}\sim 1$.
Inserting Equations (2.5)-(2.7) and (2.9) into Equation (2.8), we
have

$$\Phi^{GC}_{\gamma}(E_{\gamma})=C_{\gamma}\left(\frac{E_{\gamma}}{E_{\pi}^{GC}}\right)^{-\beta_{\gamma}}
\int_{E_{\pi}^{GC}~or~E_{\gamma}}^{E_{\pi}^{GC,max}}dE_{\pi}
\left(\frac{E_{p-p}}{E_{p-p}^{GC}}\right)^{-\beta_p}$$
$$\times N_{\pi}(E_{p-p},E_{\pi})
\frac{2}{\beta_{\pi}E_{\pi}}, \eqno(2.10)$$where the lower-limit of
the integration takes $E_{\pi}^{GC}$ (or $E_{\gamma}$) if
$E_{\gamma}\leq E_{\pi}^{GC}$ (or if $E_{\gamma}> E_{\pi}^{GC}$). In
consequence,

$$E_{\gamma}^2\Phi^{GC}_{\gamma}(E_{\gamma})=\left\{
\begin{array}{ll}
\frac{2e^bC_{\gamma}}{2\beta_p-1}(E_{\pi}^{GC})^3\left(\frac{E_{\gamma}}{E_{\pi}^{GC}}\right)^{-\beta_{\gamma}+2} \\ {\rm ~~~~~~~~~~~~~~~~~~~~~~~~if~}E_{\gamma}\leq E_{\pi}^{GC}\\\\
\frac{2e^bC_{\gamma}}{2\beta_p-1}(E_{\pi}^{GC})^3\left(\frac{E_{\gamma}}{E_{\pi}^{GC}}\right)^{-\beta_{\gamma}-2\beta_p+3}
\\ {\rm~~~~~~~~~~~~~~~~~~~~~~~~ if~} E_{\gamma}>E_{\pi}^{GC}
\end{array} \right. .\eqno(2.11)$$ Interestingly, this solution shows a typical broken power-law.

    We add the contributions of $\gamma\rightarrow e^++e^-$ to Equation (2.8) and get the following positron spectrum

$$\Phi_{e^+}(E_{e^+})=\Phi^0_{e^+}(E_{e^+})+\Phi^{GC}_{e^+}(E_{e^+}),\eqno(2.12)$$
$\Phi^0_{e^+}$ is the background flux of positron, and
$$\Phi_{e^+}^{GC}(E_{e^+})=C_{e^+}\left(\frac{E_{e^+}}{E_{\pi}^{GC}}\right)^{-\beta_{e^+}}
\int_{E_{e^+}}dE_{\gamma}\left(\frac{E_{\gamma}}{E_{\pi}^{GC}}\right)^{-\beta_{\gamma}}$$
$$\times \int_{E_{\pi}^{min}}^{E_{\pi}^{max}}
dE_{\pi}\left(\frac{E_{p-p}}{E_{p-p}^{GC}}\right)^{-\beta_p}
N_{\pi}(E_{p-p},E_{\pi})$$
$$\times \frac{d\omega_{\pi-\gamma}(E_{\pi},E_{\gamma})}{dE_{\gamma}}
\frac{d\omega_{\gamma-e}(E_{\gamma}, E_{e^+})}{dE_{e^+}}$$
$$=C_{e^+}\left(\frac{E_{e^+}}{E_{\pi}^{GC}}\right)^{-\beta_{e^+}}
\int_{E_{e^+}}\frac{dE_{\gamma}}{E_{\gamma}}
\left(\frac{E_{\gamma}}{E_{\pi}^{GC}}\right)^{-\beta_{\gamma}}$$
$$\times \int_{E_{\pi}^{GC}~or~E_{\gamma}}^{E_{\pi}^{max}}
dE_{\pi}\left(\frac{E_{p-p}}{E_{p-p}^{GC}}\right)^{-\beta_p}N_{\pi}(E_{p-p},E_{\pi})
\frac{2}{\beta_{\pi}E_{\pi}}$$
$$=\left\{
\begin{array}{ll}
\frac{2e^bC_{e^+}}{2\beta_p-1}E_{\pi}^{GC}\left(\frac{E_{e^+}}{E_{\pi}^{GC}}\right)^{-\beta_{e^+}}\left[\frac{1}{\beta_{\gamma}}
\left(\frac{E_{e^+}}{E_{\pi}^{GC}}\right)^{-\beta_{\gamma}}+
(\frac{1}{\beta_{\gamma}+2\beta_p-1}-\frac{1}{\beta_{\gamma}})\right]
\\ {\rm~~~~~~~~~~~~~~~~~~~~~~~~~~~~~~~~~ if~} E_{e^+}\leq E_{\pi}^{GC}\\\\
\frac{2e^bC_{e^+}}{(2\beta_p-1)(\beta_{\gamma}+2\beta_p-1)}
(E_{\pi}^{GC})\left(\frac{
E_{e^+}}{E_{\pi}^{GC}}\right)^{-\beta_{e^+}-\beta_{\gamma}-2\beta_p+1}
\\ {\rm ~~~~~~~~~~~~~~~~~~~~~~~~~~~~~~~~~if~} E_{e^+}>E_{\pi}^{GC}
\end{array} \right. .\eqno(2.13)$$Note that after taking average over all possible directions, the
energy of pair-creation is uniformly distributed from zero to
maximum value, i.e.,
$$\frac{d\omega_{\gamma-e}(E_{\gamma},
E_{e^+})}{dE_{e^+}}=\frac{1}{E_{\gamma}}.\eqno(2.14)$$ Equation
(2.13) shows an excess near $E_{\pi}^{GC}$ in the positron spectrum.

\section{The evidence for the GC source}

     The measurements of CR positron flux by AMS (Aguilar et al.2019) shows a
significant excess peaked at $\sim 300~ GeV$ (or $\sim 400~GeV$ if
subtracting the contribution of the diffuse background). We try to
understand it using the GC-model.

    According to the GC-model Equations. (2.11) and (2.13), if the excess of positron
spectrum at 400 GeV is arisen by the GC-effect, we predict: (i) one
can observe gamma-ray emission originating from a same GC-source in
galaxy, which has a characteristic spectrum with a broken power-law
on the two sides of $E_{\gamma}\simeq 400~GeV$; (ii) the spectral
indexes $\beta_p$ and $\beta_{\gamma}$ of this gamma-ray spectrum
are same as that of the positron spectrum; (iii) the value of
$C_{\gamma}$ for this GC-source is larger than that of other similar
gamma-ray sources since it is closest to the earth; (iv) this
GC-source has a relatively large value of $\beta_{\gamma}$ since it
implies a strong $\gamma\rightarrow e^++e^-$ conversion mechanism.

 \begin{figure}[htb]
    \centering {
        \includegraphics[width=0.96\columnwidth,angle=0]{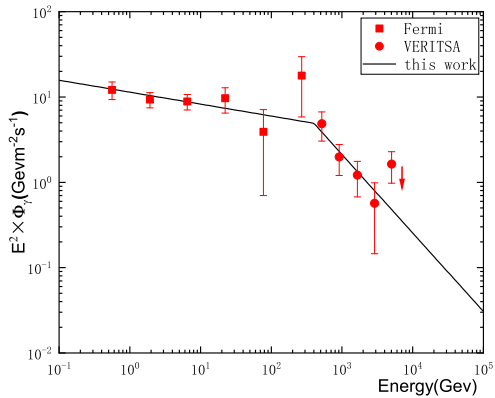}
    } \caption{\label{fig:fig1} The predicted cosmic gamma-ray spectrum by the
    GC-model Equation (2.11) (the broken lines) and comparisons with the
    Fermi-LAT and VERITAS data for Tycho (Archambault et al.2017). The last point at 5 TeV is
    excluded since its complete uncertainty. The parameters are
    $E_{\pi}^{GC}=400~GeV$, $\beta_p^{Tycho}=0.89$,
    $\beta_{\gamma}^{Tycho}=2.14$ and $C_{\gamma}^{Tycho}=6\times
    10^{-10}~GeVm^{-2}s^{-1}$. }
\end{figure}

\begin{figure}[htb]
    \centering {
        \includegraphics[width=0.96\columnwidth,angle=0]{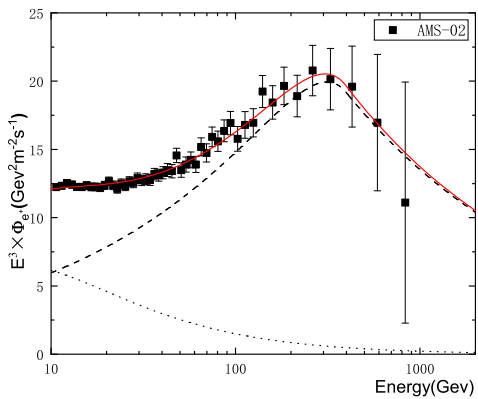}
    } \caption{\label{fig:fig2}Using the GC-model Equation (2.13) (a dashed curve) to fit the positron flux
    (Field et al.1977). The contributions of background (a point curve)
    are taken from a diffuse model (Aguilar et al.2013) but the parameters are adjusted. The
    solid curve is a total spectrum, its $\chi^2/d.o.f.=19.65/44$
    between 10 and $10^3$ GeV with the free parameters
    $C_{e^+}=4.25\times 10^{-10}, \beta_{e^+}=0.460,
    \beta_{\gamma}^{e^+}=2.14, \beta_p^{e^+}=0.89, C_d=0.067,
    \gamma_d=-3.86, E_1=6.33, E_0=1.13$ and $
    E^{GC}_{\pi}=400~GeV$. }
\end{figure}

\begin{figure}[htb]
    \centering {
        \includegraphics[width=0.96\columnwidth,angle=0]{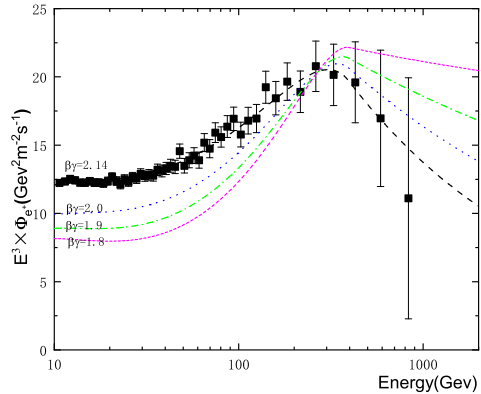}
    } \caption{\label{fig:fig3}The contributions of the GC-source to the positron flux with
    different values of $\beta_{\gamma}$. Where the contributions of the background flux are added.}
\end{figure}

\begin{figure}[htb]
    \centering {
        \includegraphics[width=0.96\columnwidth,angle=0]{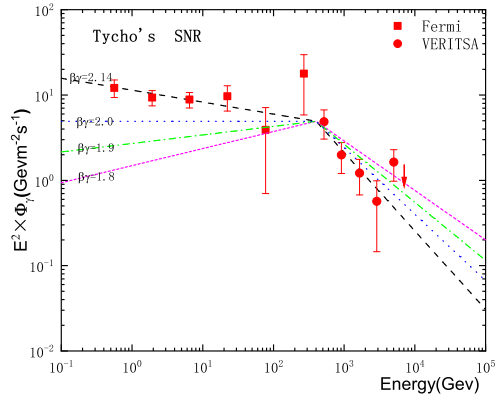}
    } \caption{\label{fig:fig4}The contributions of the GC-source to the Tycho's gamma-ray flux with
    different values of $\beta_{\gamma}$.}
\end{figure}

    Interestingly, supernova remnant (SNR) G120.1+1.4 is such a possible GC-source.
In history it was observed by Tycho in 1572 and hereafter referred
to as Tycho. An updated gamma-ray spectrum of Tycho's SNR by VERITAS
and Fermi-LAT shows a clear broken power-law. Although some other
models may fit the updated gamma-ray spectra after adjusting their
parameters (Archambault et al.2017; Bykov et al. (2018);Cristofari
et al.(2018)), as a new attempt, we try to use the GC-effect
(Equation (2.11)) to fit the Tycho's spectrum. Figure 1 gives our
fitting gamma-ray spectrum and the comparison with the Tycho's
spectrum, where the GC-threshold $E_{\pi}^{GC}=400~GeV$,
$\beta_p^{Tycho}=0.89$, $\beta_{\gamma}^{Tycho}=2.14$ and
$C_{\gamma}^{Tycho}=6\times 10^{-10}~GeVm^{-2}s^{-1}$ are fixed by
the VERITAS and Fermi-LAT data. We get a reasonable fitting quality
criteria

$$\chi^2/d.o.f.=\sum_i\frac{(\Phi^{pre}_i-\Phi_i^{obs})^2}{\sigma^2_{sys,i}+\sigma^2_{stat,i}}
\times \frac{1}{d.o.f.}=\frac{3.24}{6}=0.54. \eqno(3.1)$$

    The one point at the high energy end of the VERITAS data is the most
discrepant point relative to the fit. We noticed that the
GC-threshold is nuclear mass $A$-dependent (see the next section). A
rough estimation shows that the GC-effect in $p-p$ collision may
contribute an excess to the gamma and electron-positron spectra
above $20~TeV$ (Zhu et al. 2018). We wait for future more sensitive
measurements.

        Now we assume that the excess of the positron spectrum recorded by AMS
arises from Tycho's SNR. One can write the total positron flux,
which is the sum of the source term Equation (2.13) and a diffuse
background term (Aguilar et al.2013)

    $$\Phi^{0}_{e^+}(E)=C_d\frac{E^2}{\hat{E}^2}\left(\frac{\hat{E}}{E_1}\right)^{\gamma_d},
\eqno(3.2)$$ where $\hat{E}=E+E_0$. We substitute the fixed
parameters $E_{\pi}^{GC}=400~GeV$, $\beta_p^{e^+}=0.89$ and
$\beta_{\gamma}^{e^+}=2.14$ into Equation (2.13) and adjust
$\beta_{e^+}$, $C_{e^+}$ and the remaining parameters in Equation
(3.2). Note that $\beta_p^{e^+}=\beta_p^{Tycho}$ and
$\beta_{\gamma}^{e^+}=\beta_{\gamma}^{Tycho}$ are fixed. The result
is presented in Figure 2. The fitting $\chi^2/d.o.f.=19.65/44=0.45$.

\section{Discussions and summary}

   The conclusions of this work are based on Equations (2.1) and
(2.2)under the GC-condition. This is a plausible but unproved
assumption. Although the data from a range of supernova remnants
including Tycho's can be approximately described by several models
including leptonic and hadronic components without the GC-effect.
However, the GC-effect in hadronic framework can connect two kinds
of astrophysical observations perfectly, namely the positron excess
in AMS and the broken power-law for Tycho's gamma ray spectrum. This
feature is what we want to emphasize.

    For illustrating the sensitivity of the fitting quality
to the selection of the parameters, we draw the contributions of the
GC-source to the positron- and gamma-ray fluxes with different
values of $\beta_{\gamma}$ in Figures 3 and 4, respectively. The
sensitivity of the results to the parameter selection shows that our
$\chi^2/d.o.f.\simeq 0.5$ is not an accidental coincidence. It seems
that the high energy data points of AMS do not place significant
constraints against the value of $\beta_{\gamma}$, but rather it is
the low-energy Fermi-Lat points that constrain $\beta_{\gamma}$.
However, considering the correlation between the low-energy point
and the high-energy point, the restriction of Figures 3 and 4 on the
parameter $\beta_{\gamma}$ is strict.

    The results in Section 3 are consistent with our predictions: (i)
$E_{\pi}^{GC}=400~GeV$; (ii) $\beta_p^{Tycho}=\beta_p^{e^+}$ and
$\beta_{\gamma}^{Tycho}=\beta_{\gamma}^{e^+}$.

        Finally, we should explain why no GC-signals are observed at the
LHC. The GC-threshold $E_{\pi}^{GC}$ is target $A$-dependent, since
the nonlinear term of the QCD evolution equation should be re-scaled
by $A^{1/3}$ and $E_{\pi}^{GC}$ decreases with increasing A (zhu
2017). However, A-dependence of $E_{p-A}^{GC}$ is a complicated
problem, which relates to the distribution and structure of the
GC-source. We have not available input distributions of the
nonlinear QCD evolution to precisely predict the GC-threshold.
According to a roughly estimation in (Zhu et al. 2017), we have
$E_{p-p}^{GC}>E_{p-A}^{GC}>E_{A-A}^{GC}$. Using Equations
(2.1)-(2.7), the center-of-mass energy corresponding to
$E_{\pi}^{GC}=400~GeV$ is

$$\sqrt{S^{GC}_{A-A}}\simeq\sqrt{2m_pE_{A-A}^{GC}}=\sqrt{2m_p}e^{b-a}E_{\pi}^{GC}=5.5~TeV.\eqno(4.1)$$
The ALICE and ATLAS collaborations at the LHC have measured Pb-Pb
(and p-Pb) collisions till $\sqrt{S_{Pb-Pb}}=5.02~TeV$ (and
$\sqrt{S_{p-Pb}}=8.16~TeV$) (Rode et al.2019; ATLAS Collaboration
2019).  We consider that $\sqrt{S_{Pb-Pb}^{GC}}>5~TeV$ and
$\sqrt{S_{p-Pb}^{GC}}>8~TeV$, it implies that the GC-effect is
entering (or will enter into) a measurable energy range.

    In summary, a QCD research predicts that gluons in protons may condensed at a
critical momentum in high energy collision of proton-proton. We use
the GC-model to perfectly explain two seemingly completely different
events of CR spectra: an excess in positron flux and the broken
power-law in gamma-ray spectrum of Tycho. We consider that the
excess in the CR positron spectrum observed by AMS originates mainly
from the GC-effect in Tycho's supernova remnant.

\noindent {\bf ACKNOWLEDGMENTS} We would like to thank the anonymous
reviewer for his/her insightful comments on the manuscript. The
comments have really helped us to improve our manuscript to the best
possible extent. This work is supported by the National Natural
Science of China (No.11851303).

\end{document}